\documentclass[epj]{svjour}
\usepackage{graphicx}
\usepackage{textgreek}
\begin{document}
%
\title{Review and outlook of accelerator-related codes and their interplay with the experiments software}

\author{Manuela Boscolo\inst{1,2} \and  Helmut Burkhardt\inst{2} \and  Gerardo Ganis\inst{2} \and Cl\'ement Helsens\inst{2} 
}                     
\offprints{}          
\institute{INFN \and CERN}
\authorrunning{Manuela Boscolo et al.}
\titlerunning{Accelerator-related codes and their interplay with the experiment's software}
\date{{\it (Submitted to EPJ+ special issue: A future Higgs and Electroweak factory (FCC): Challenges towards
discovery, Focus on FCC-ee)}}
%
\abstract{
Powerful flexible computer codes are essential for the design and optimisation of accelerator and experiments. 
We briefly review what already exists and what is needed in terms of accelerator codes.
For the FCC-ee it will be important to include the effects of beamstrahlung and beam-beam interaction as well as machine imperfections and sources of beam induced backgrounds relevant for the experiments
and consider the possibility of beam-polarisation.
The experiment software Key4hep, which aims to provide a common software stack for future experiments, is described and the possibility of extending this concept to machine codes is discussed.
We analyse how to interface and connect the accelerator and experiment codes in an efficient and flexible way
for optimisation of the FCC-ee interaction region design, and discuss the possibility of using shared data formats as an interface.
} 
\maketitle

\section{Introduction}
\label{intro}
The international Future Circular Collider (FCC) study aims to design p-p, $\rm e^{+}e^{-}$ and e-p colliders to be built in a new 100~km tunnel in the Geneva region. 
The $\rm e^{+}e^{-}$ collider (FCC-ee) has a centre of mass energy range between 91.2 and 365~GeV with instantaneous luminosities as high as $\rm 2.3\times10^{36}cm^{-2}s^{-1}$ and $\rm 1.55\times10^{34}cm^{-2}s^{-1}$, respectively~\cite{FCCCDRLept}. The design of the interaction region is crucial to reach such unprecedented energies and luminosities. 

\begin{figure}[ht!]
\centering\includegraphics[width=0.5\textwidth]{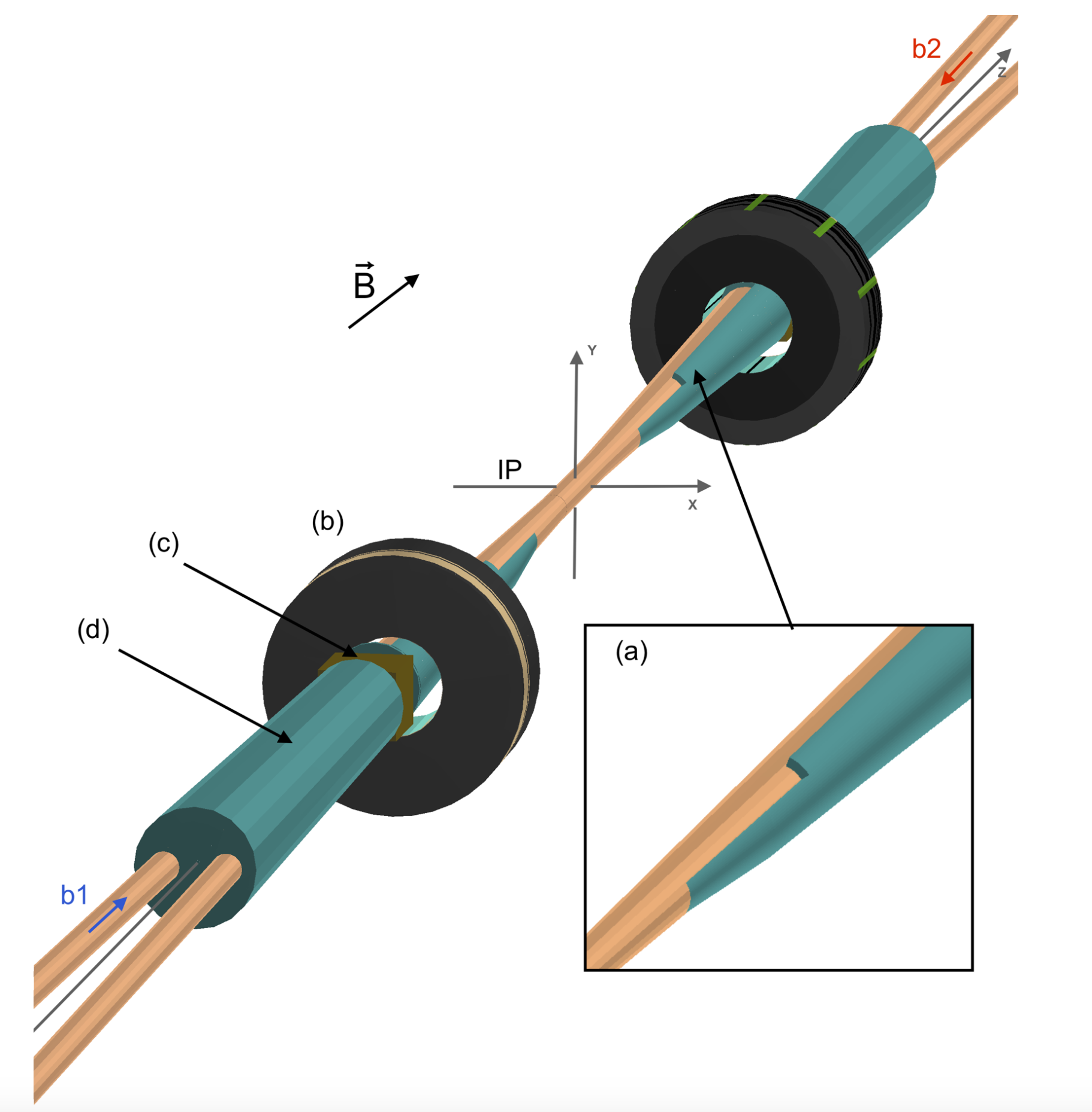}
\caption{Sketch of the FCC-ee interaction region. (a) Detail of the shielding; (b) luminometer; (c) HOM absorber; (d) thick tungsten shielding\,\cite{FCCCDRLept}.}
\label{GEANT4_IR}
\end{figure}

The main characteristics of the interaction region optics design
is determined by the crab-waist scheme with a local chromatic correction system and a horizontal crossing angle of
30~mrad at the interaction point.
A description of the main challenges of the interaction region and machine detector interface (MDI) design can be found in Ref.~\cite{ref:mdi_epj}.
The baseline optics for the FCC-ee double-ring collider is described in Ref.~\cite{ref:prab-oide}.
 The total synchrotron radiation power is limited by design at 100\,MW for the two beams and consequently the stored current per beam 
varies from 1.4\,A at Z to 5.4\,mA at the $\mathrm{t \bar{t}}$ data taking stage.
Following the LEP-2 experience where the highest local critical energy was 72\,keV for photons emitted at 260\,m from the IP~\cite{gvh}
 the FCC-ee optics design maintains critical energies from bending magnets below 100\,keV
starting at 100\,m from the interaction point; the critical energy from the first bend after the interaction point 
is higher, at 691\,keV for the $\mathrm{t \bar{t}}$ threshold. 
 An asymmetric optics has been designed to meet these critical energy goals in the interaction region.
Synchrotron radiation mask tips 
are placed in the horizontal plane just in front of the first final focus quadrupole at  2.1\,m  from the interaction point.
 The free length between the interaction point and the first final focus quadrupole  is 2.2\,m, which is inside the detector.
Figure~\ref{GEANT4_IR} shows the {\sc geant4} model with the shielding and the luminometer that was used for background simulation studies~\cite{FCCCDRLept}.
The  compactness of the MDI design, determined by the space available to host all the necessary components like the first final focus quadrupole and the anti-solenoids being placed inside the detector, poses interesting technical challenges. 
Part of the challenge is the development of modern flexible software tools 
for the beam optics model, including the beam induced background scattering processes with an interface with the experiments.\par
This essay is organised as follows.
We give an overview of the existing accelerator codes related to the interaction region and MDI design in Section~\ref{sec:2}; in Section~\ref{sec:fccsw} we describe the experiment software and the current strategy to interface the accelerator and experiment codes; in Section~\ref{sec:geometry} we discuss the geometrical description of the relevant elements. Finally, in Section~\ref{sec:concl} we summarise the status and the challenges ahead.

\section{Review of the accelerator codes and MDI considerations}
\label{sec:2}

From the beginning of the studies for FCC, it was realised that the design of the interaction regions is particularly challenging and it is necessary that the requirements of the machine and experiments should be evaluated and optimised concurrently.

\subsection{Lattice codes}
\label{sec:lattice}
The most popular code used for accelerator design at CERN and several other laboratories is the Methodical Accelerator Design program ({\sc mad}).
The {\sc mad} program has been developed, maintained and upgraded for more than 3 decades. The main version used at LEP was {\sc mad8}\,\cite{Grote:1991zp}, written in FORTRAN V. Since then, it was largely rewritten as {\sc mad-x} using a combination of FORTRAN95 and C, and later also C++ for selected modules. The activity started in the new millenium\,\cite{Grote:2003ct}, coinciding with the end of LEP operation and the shift of priorities motivated by developments for the LHC and its injectors, where synchrotron radiation only plays a minor role.

For the FCC-ee design, we also profit from the more recent experience and code developments for lepton colliders, by working with the Strategic Accelerator Design ({\sc sad})\,\cite{sad} program used for SuperKEKB. {\sc sad}, like {\sc mad-x}, is an accelerator lattice design code, developed independently at KEK and therefore it provides a good opportunity for comparison and cross-checking.
Both {\sc sad} and {\sc mad-x} read the machine description from formatted text files, using their own specific commands to specify magnet types, strengths and position. These input files are typically referred to as sequence files.
A translator to convert the {\sc sad} lattices into {\sc mad-x} format exists.

The {\sc mad-x} lattice input text format allows the specification of aperture information like beam pipe shapes, sizes, and more recently also information on materials, provided as formatted comments\,\cite{Deniau:2019nca}.
Even for large machines like FCC with the order of 10\,000 magnets, the size of the sequence files is always manageable, often less than a megabyte.
The basis for the MDI work is the lattice description in {\sc mad-x} format, typically starting from {\sc sad} translated to {\sc mad-x} format, with additional information on aperture and beam pipe material.

The output from {\sc mad-x} provides extra information per element like, beam position, beam size, TWISS parameters (beta functions, phase advance ..), $6\times 6$ element transfer matrices and optionally $6\times 6\times 6$ second order transfer maps. The {\sc mad-x} output format is known as TFS-format\,\cite{Grote:1991zp}, a human readable tabular format with extra general header information specifying parameters that apply to the whole machine like nominal beam energy and beam particle type.
Even if we select all options and write numbers with 17 digits to avoid any loss in precision, the file sizes  remain below 100 megabytes.

\begin{figure}[ht!]
\centering\includegraphics[width=0.9\textwidth]{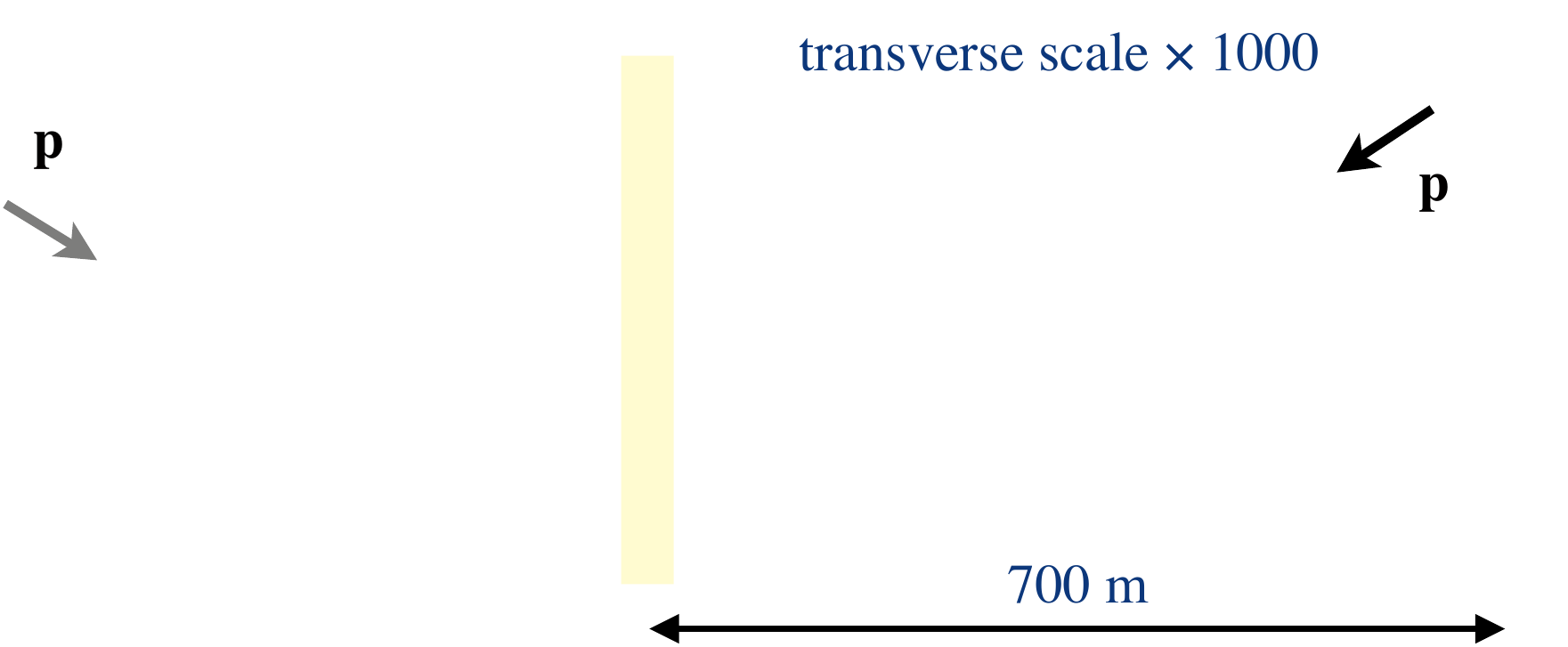}
\caption{Illustration of the FCC-hh interaction region using {\sc root}, and interfacing the {\sc mad-x} generated machine layout with 
{\sc MDISim} to {\sc geant4} to track the protons through the interaction region and generate synchrotron radiation photons~\cite{Collamati_2017,Collamati:2017hlr}.}
\label{FCC_MADX_MDISim_ROOT_GEANT4}
\end{figure}

\subsection{MDI considerations}
\label{sec:mdi}
The software used in MDI studies must be able to simulate in detail what happens in the interaction region.
The beam pipe aperture should be sufficiently large to allow the beam particles to pass without any major losses for all modes of operation, including injection, and also be compatible with possible failure scenarios. 
Heating by HOM (higher order modes losses, induced by the electromagnetic fields of the intense bunched beams) must be minimised.
Even under optimal conditions, the intense particle beams stored in the FCC will always result in some level of particle losses and synchrotron radiation, the latter is considered by the experiments as unwanted beam induced backgrounds. The modeling of the beam induced background effects has often been performed with custom Monte Carlo codes and relying on input and geometry data in specific formats which have to be written or modified by hand.

These days, computers are sufficiently powerful and have enough memory and storage capacity
that it could be possible to build a supercode that combines all that is needed in a single program.
Another possibility would be to make all codes available from a unified code library.
This was already attempted in the nineties\,\cite{Iselin:1996jz} for accelerator codes, but with rather limited adoption and lack of support.
The choices made can also be influenced by sociological perspectives and personal preferences. Working with large programs which have grown historically can be intimidating and may appear less rewarding than creating well defined smaller programs and code pieces associated with the names of a few authors.

For FCC (ee and hh), we have made an effort in the development of {\sc MDISim}~\cite{ref:mdisim} to use and combine existing codes as much as possible using a light, flexible interface, minimising the need for hand coded geometries, and privileging open source and well supported codes and exchange formats. We use {\sc MDISim} to read the TFS files generated by {\sc mad-x} and automatically translate them into an exchange format, directly readable by {\sc geant4}~\cite{GEANT4} and {\sc root}~\cite{ROOT}. The format used is GDML\,\cite{GDML} for the geometry information, complemented by magnet strengths, initial beam positions and directions provided by automatically generated human readable text files. {\sc geant4} is used to track the particles through the interaction region with generation of secondary particles and simulation of interactions in materials. The ROOT Event Visualization Environment is used for the display.
An example is shown in Figure~\ref{FCC_MADX_MDISim_ROOT_GEANT4} \,\cite{Collamati_2017}.

Further information on the geometry and material outside the beam pipe like vacuum equipment, shielding, magnet material or the experiment detectors can then be added on the GDML level.
Many commercial and open software codes are being developed that can work with GDML and interface with other flexible geometry descriptions including CAD formats\,\cite{instep,salome,cadmesh,blender,step,vtcad,sw2gdml,cadmc,pyg4ometry}.

\subsection{Beam induced backgrounds}
\label{sec:beambkg}
To simulate and minimise the impact of all relevant processes that can result in the loss of beam particles or secondary particles generated by the beams in the experiment detectors is a complex task. Beam-gas, Touschek and thermal photon scattering as well as synchrotron radiation will always be present, even if beams are not colliding, and they require the whole ring to be studied.
For FCC-ee, the minimisation of synchrotron radiation effects is of primary importance and has strongly influenced the basic design and layout choices.
The collisions of the beams in the interaction regions will generate additional losses
and produce synchrotron radiation by deflection in the electromagnetic fields of the opposing beam (referred to as beamstrahlung).

The {\sc MDISim} code is capable of efficiently generating the accelerator and beam pipe geometry for shower simulations for the whole ring. The {\sc geant4} toolkit\,\cite{GEANT4} has code for all major particle scattering processes including the generation of synchrotron radiation\,\cite{Burkhardt:2007zza}, as well as tracking in magnetic fields, and is well suited for detector simulations. {\sc geant4} can  be considered as a candidate for a supercode that simulates both machine and experiment, and has been used as the basis for the combined codes {\sc g4beamline} and {\sc bdsim}\,\cite{G4beamline,BDSIM}.

For benchmarking purposes, we use {\sc geant4} directly to track a few particles over several turns in small machines, and have been contributing to developments to improve the tracking precision in {\sc geant4}. For large machines like FCC this is not realistic at present:
we would need to consider some $10^{11}$ particles per bunch, circulating many times in a 100\,km ring, to be able to determine the effects of radiation and the loss of a tiny fraction of the circulating particles in the interaction regions.
At present, we restrict the {\sc geant4} based simulations to roughly a kilometer around the interaction region and work with 
other, more dedicated codes to complete these studies when needed. 
To gain speed, it is easier to guarantee a high level of numerical precision in accelerator codes that track deviations from the design path rather than tracking absolute positions.

The synchrotron radiation is emitted in excellent approximation in beam direction and not deflected by magnet fields. With scattering and reflection processes included, we find that only the synchrotron radiation generated in a limited range (some hundred meters) around the interaction is relevant as a source of detector backgrounds, and that more general effects like non-Gaussian tail generation in collisions can be taken into account by a proper choice of the beam distribution. Details of the {\sc geant4} based synchrotron radiation background simulations are described 
in Ref.~\cite{MarianThesis}. We also make comparisons with the more dedicated {\sc sync\_bkg}~\cite{ref:sync} and {\sc synrad+}~\cite{ref:synrad} codes.
Non-gaussian tails were observed at LEP\,\cite{Burkhardt:1999xb}. They can be generated by scattering processes, and be enhanced by non-linearities in the beam-beam interaction or strong sextupoles in combination with machine imperfections.  A popular code, much used at LHC to study the effect of non-linearities and imperfections in multiturn tracking is {\sc SixTrack}\,\cite{SixTrack}.
It would have to be adapted for $\mathrm{e^+e^-}$ multiturn tracking, for example to account for the synchrotron radiation damping. A new tracking tool named {\sc Xtrack}, part of the Xsuite project\,\cite{xsuite} is being considered for collimation studies at FCC-ee.

The {\sc guineapig}\,\cite{guineapig} code is used as generator of beamstrahlung and radiative Bhabha scattering in the interaction regions. We also use the {\sc bbbrem}~\cite{Kleiss:1994wq} code as generator for the simulation of radiative Bhabha scattering . This process is characterised by an energy loss of one of the colliding particles and very small scattering angles, such that the scattered particles remain within the beam pipe.
The process has been integrated into {\sc sad} and was in particular used for simulations of FCC-ee at the Z energy, where the beam intensity and luminosity are highest.

Particle losses by beam-gas and thermal photon scattering in multiple turns around the ring, are taken into account using {\sc sad, mad-x}, or {\sc ptc} for the transport element by element, performing aperture checks at element boundaries. The results\,\cite{ciarma} were compared
with the more detailed {\sc geant4} simulations performed around the interaction regions~\cite{Boscolo:2018ltl} and they were found to be in good agreement.
The vacuum pressure profile in the MDI area, as well as in the whole ring, can be used as input for the beam-gas scattering simulations rather than assuming a constant pressure profile inside the beam pipe.
 The local pressure profile can be evaluated by means of the Monte Carlo code {\sc molflow+}\,\cite{molflow}. 
{\sc molflow+}  provides detailed 3D calculations of vacuum properties in the molecular flow regime, such as pressure profiles, effective pumping speeds, and adsorption distributions which are of interest mainly for vacuum engineers. 
It also allows the simulation of gas propagation in CAD imported geometries and simulates pumpdown processes.

Touschek scattering is an intra-beam scattering particularly relevant for low energy storage rings,
and is
the major beam lifetime limitation for lepton colliders like DA\textPhi NE, SuperKEKB and all the modern low and ultra-low emittance light sources.
For the high energy FCC-ee, a simpler, more dedicated Monte Carlo embedded in particle tracking as developed for DA\textPhi NE~\cite{Boscolo:2012ws} and used for SuperKEKB should be fully sufficient.

Beam polarisation will be important for the FCC-ee and in particular for the precise determination of the beam energy.
It requires studies of the whole ring including realistic modeling of imperfections and optics corrections~\cite{Gianfelice-Wendt:2016jgk}.
A good candidate for detailed simulations with polarisation is the {\sc sitros} code~\cite{Kewisch:1983uq}.
Another good candidate is {\sc bmad}~\cite{bmad} which is a flexible software toolkit for the simulation of charged particles and X-rays including the spin tune and polarisation.

\section{Experiment software}
\label{sec:fccsw}

The software used to study the physics potential of FCC goes generically under the name of FCCSW and comprises a set of tools covering the needs of an experiment: signal generation, detector description and simulation, event reconstruction and final analysis~\cite{fccswchep}. 
FCCSW is a result of a process started just after the FCC project kick-off in 2014. The design goal has been to support physics and detector studies with parameterised, fast, and full simulation, also allowing a mixture of the three. It has to be modular enough to allow for evolution, allowing component parts to be improved separately. Finally it has to allow multi-paradigms for analysis, with C++ and Python at the same level. The strategy to meet these challenging requirements has been to adopt solutions developed for LHC, such as the Gaudi framework~\cite{gaudi} and to look at ongoing common projects; among the latter, it is worth mentioning those developed under the AIDA EU R\&D effort~\cite{AIDA}: {\sc podio}~\cite{podio}, used to define the event data model, and {\sc dd4hep}~\cite{dd4hep}, used for the geometrical description of all the elements relevant for the physics measurements, i.e.\ the sensitive and passive elements of the sub-detectors, supports, magnet and elements of the interaction region affecting the detector performance, such as the beam pipe and other elements which can scatter or produce particle debris in the detector.

The Gaudi framework implements an architecture in which data flow through a transient data store (in memory), where they can be modified by algorithms representing the various steps of the data processing chain, e.g. generation, simulation or reconstruction. Readers and converters are special algorithms that can inject data into the transient store for processing.   

Because of its capability of coping effectively with the data processing needs of High Energy Physics (HEP) experiments and the challenges of HL-LHC, Gaudi has also been chosen as the main framework of the Key4hep common software project, around which FCCSW is going to evolve in the future.

\subsection{MDI induced backgrounds}
\label{sec:fccswmdibkg}
The processes described in Section~\ref{sec:beambkg}, in addition to influencing the beam lifetime and stability, can be sources of backgrounds - and therefore of systematic effects - for the physics measurements and need to be controlled as precisely as possible.
Before FCC-ee, the use of the codes described in Section~\ref{sec:mdi} was restricted to a few experts who were producing estimates of what turned out to be small effects, which were possibly mentioned as upper limits on systematic errors in final analysis. The only beam-related effects included by physicists in the simulation of the detector response were spreads in the beam energy and the position of the effective interaction point, which are important and non-negligible but do not cover the full picture.
The unprecedented design luminosities of FCC-ee require a better evaluation of all the effects, including those of Section~\ref{sec:beambkg}, which can only be obtained by simulating these backgrounds in the experimental apparatus to properly estimate detector occupancies and the level of spurious objects, such as additional tracks.
This requires inter-operability of the relevant codes with FCCSW.

\subsection{Interplay between accelerator and experiment codes}
\label{sec:fccswmdi}
There are several levels of software programs that can inter-operate. The one that we will pursue in this case is one which is at the lowest level and which goes through common data formats and which can also work for programs running on different hardware or operating systems. We have seen in Section~\ref{sec:2} that the codes for MDI-induced backgrounds typically produce outputs in the form of formatted text files. There is no common output format for all the programs, but there is enough information to understand the outputs and use them in other contexts. 

The underlying idea is to develop a set of Gaudi readers and/or converters to inject the events produced by the MDI-background codes in the data processing chain.\footnote{While in the default running mode inter-operability is through persistent files, the availability of dedicated readers and/or converters opens the way for alternative inter-operation options, for example through FIFO channels~\cite{FIFOref}.}

FCCSW will then simulate the interaction of these particles in the detector to evaluate occupancies and levels of spurious objects.
Eventually FCCSW will provide the possibility to 
overlay these events on signal events for a more detailed background simulation, possibly with a weighted mixture of MDI processes.

\subsection{Towards an MDI "supercode"}
\label{sec:supercode}
By `supercode' we do not mean a new big program doing everything but a common interface to all relevant codes to effectively simulate a single big program behaviour. The ultimate goal is that physicists wanting to study these backgrounds are able to do so from any of the supported computing infrastructures.

A software ecosystem based on Gaudi allows an approach of this type.
The relevant components which need to be provided and/or identified are the following:
\begin{enumerate}
    \item A shared file system available on the computing infrastructures supported for the project; 
    \item A software stack providing all the relevant applications built in a coherent way;
    \item A set of good default configuration files available in the shared file system accessible by the relevant applications;
    \item A wrapper to run external applications in Gaudi;
    \item A set of Gaudi readers and converters, as mentioned in the previous section;
    \item A set of application command line controls covering all the identified needs.
\end{enumerate}
For the shared file system the obvious choice is
CernVM-FS~\cite{cvmfs,cvmfstwo} which is now ubiquitous in HEP communities and beyond. 
The Key4hep stack~\cite{key4hep} is the choice for 2, with all the stack software available under {\tt /cvmfs/sw.hsf.org/}. Some of the relevant codes, e.g. {\sc guineapig}, are already available in Key4hep. Part of the work would be to make sure that all the relevant software are in a form suitable for being added and maintained in the common stack.  
Good default configuration files should be the result of the detailed evaluation of each of the codes mentioned earlier; they could be stored on the shared CernVM-FS repository, though the possibility to use different settings should be maintained.
The integration with Gaudi, points 4 and 5, is part of the work, and no insurmountable issues are anticipated.
The identification of a common set of switches to cope with all the codes will need some iterations, though, again, it should not pose insolvable problems.

The whole integration process just described is being proof-tested with {\sc guineapig} with promising results already. Detailed documentation of this prototype work and of course of all the components described in this sub-section are part of the work. It should contain examples at different levels, together with a detailed reference guide.

\section{Aspects related to geometry description}
\label{sec:geometry}
The description of the geometry and materials of the relevant accelerator and detector elements is a crucial ingredient of most of the codes discussed in this essay. Having a coherent description, based on the same source of information is highly desirable for several reasons, not least to reduce the risk of errors due to several implementations of the same item. 

The design and implementation of these elements follows different paths depending on the nature of the element. {\it Detector components} are designed starting from a detector concept and then it is mostly space constraints that are applied. The tool chosen to describe the relevant concepts is DD4hep~\cite{dd4hep}, an open source toolset introducing the compact detector description concept, provided through minimalistic XML formats, to allow composition of basic sub-detector elements to form complex detector structures.
DD4hep was developed for conceptual design studies and initially applied to linear collider cases. However, its flexibility and generality quickly appealed the LHC community; as of today, DD4hep has been adopted by CMS for use starting from Run3 and is being seriously considered by LHCb and ATLAS.
{\it Accelerator elements} usually arise from optimisation studies carried on with CAD engineering tools which allow working in 3D with the solid features which are essential for the task. Currently the tool mostly used is {\sc Autodesk Inventor}~\cite{adeskinv}, though {\sc catia}~\cite{catia} is currently being evaluated; both these tools are commercial, because the open source CAD tools currently available do not provide the required quality. 

While there are good reasons for both approaches, the desirable requirement to have the same source of information becomes a challenge, because a satisfactory conversion procedure of CAD supported formats to DD4hep is currently missing.
DD4hep has some capabilities to read CAD formats though an interface to the external open source library {\sc AssImp}~\cite{assimp}. As well as the fact that only read support is currently implemented - which would result in one-way only conversion, namely CAD to DD4hep - the major limitations come from the CAD file formats currently supported by {\sc AssImp} and the overlap with those supported by {\sc Autodesk Inventor}. Early investigations have shown that currently the only testable solution is to use the STL (Standard Tessellation Language) format~\cite{stlref}. However, this format has some inherent limitations, since it focuses on surfaces and does not seem to provide a natural way to describe material information, which is a must-have for any simulation activity.
Other conversion options are under investigation such as the ones discussed in Ref.~\cite{pyg4ometry} or the suggestions available for {\sc Geant4} users~\cite{instep,salome,cadmesh,blender,step,vtcad,sw2gdml,cadmc,pyg4ometry}. Finding a satisfactory solution to the mutual conversion between CAD formats and DD4hep is clearly one of the challenges of the MDI detector studies.

\section{Conclusion}
\label{sec:concl}
In this essay we reviewed the main aspects affecting accelerator codes and their interplay with experimental software.
The existing accelerator codes are the result of many years of development and have been validated with respect to various accelerator facilities. Often two or more alternative codes for each of the different operational and experimental aspects exist. The challenge in these cases is to get full control of the codes, often not available in version repositories. 
We need to facilitate access and configuration and, when relevant, provide clear and, if required, solid ways of combining the results (see, e.g., the case of synchrotron radiation).

For the integration with experimental software, the main challenge is to provide the relevant Gaudi components to enable the interplay between accelerators codes and the data processing chain.
Finally, to achieve the objective of a single geometry source for all the components, a solid conversion solution to and from CAD will be required.

The interaction region design for FCC is particularly challenging and requires a combined optimisation of accelerator, engineering and experiment aspects.
A comprehensive and flexible software environment as outlined in this essay will be very helpful for the MDI design of a future collider.



%
%
%

%
\bibliographystyle{ieeetr}
%

\end{document}